\documentclass[twocolumn]{aastex631}

\shorttitle{
Radio images inside of relativistic jets
}
\shortauthors{Ogihara, Kawashima \& Ohsuga}

\begin{document}

\title{
Radio Images inside Highly Magnetized Jet Funnels Based on  Semi-Analytic GRMHD Models
}

\correspondingauthor{Taiki Ogihara}
\email{taiki.ogihara@gmail.com}

\author{Taiki Ogihara}
\affiliation{Center for Computational Sciences, University of Tsukuba, Ten-nodai, 1-1-1 Tsukuba, Ibaraki 305-8577, Japan}

\author[0000-0001-8527-0496]{Tomohisa Kawashima}

\affiliation{Institute for Cosmic Ray Research, The University of Tokyo, 5-1-5 Kashiwanoha, Kashiwa, Chiba 277-8582, Japan}

\author[0000-0002-2309-3639]{Ken Ohsuga}
\affiliation{Center for Computational Sciences, University of Tsukuba, Ten-nodai, 1-1-1 Tsukuba, Ibaraki 305-8577, Japan}

\begin{abstract}
{
By performing general relativistic radiative transfer calculations, we show the radio images of relativistic jets including highly magnetized regions inside jet funnels, based on steady, axisymmetric, and semi-analytic general relativistic magnetohydrodynamics models.
It is found that multiple ring images appear at the photon frequency of 230 GHz for nearly pole-on observers, because of the strong light bending effect on photons generated at the separation surfaces which is the boundary between the inflow and outflow flows in the jet funnel.
A bright teardrop-shaped component, which extends from the bright rings of the separation surface, also appears in the counter jet region. 
The diameter of the brightest outermost ring originated from the counter jet is $\sim 60 ~ \mu {\rm as}$, which is consistent with the ring-like images of M87 at 86 GHz observed with GMVA, ALMA and GLT, whose ring-diameter is $\sim 64^{+4}_{-8} ~\mu {\rm as}$.
The thinner and smaller-diameter rings are exhibited when the black-hole spin magnitude is higher.
These morphological features are expected to appear without being prominently affected by the detailed MHD-plasma parameters of our GRMHD jet model, since the location of the separation surfaces is mainly regulated by the black hole spin.
Our GRMHD model and the emission features of the images in the horizon-scale, highly magnetized jet funnel may be tested by future observations, e.g., the next-generation Event Horizon Telescope and the Black Hole Explorer. 
}
\end{abstract}


\section{Introduction} \label{sec:intro}
Relativistic jets have been observed in active galactic nuclei (AGNs).
High-resolution very long baseline interferometry (VLBI) radio observations have resolved the detailed emission structures of AGN jets.
Limb-brightened structures have been observed at millimeter-centimeter wavelengths in jets of M87 \citep{Kovalev2007The-Inner-Jet-o, Walker2008A-VLBA-movie-of, Walker2018The-Structure-a, Hada2011An-origin-of-th, Hada2016High-sensitivit, Mertens2016Kinematics-of-t, Kim2018The-limb-bright}, Mrk 501 \citep{Piner2009Significant-Lim}, Mrk 421 \citep{Piner2010The-Jets-of-TeV}, Cyg A \citep{Boccardi2016First-3-mm-VLBI}, 3C 84 \citep{Nagai2014Limb-brightened,Giovannini2018A-wide-and-coll}, Cen A \citep{Janssen2021Event-Horizon-T}, 3C273 \citep{Bruni2021RadioAstron-rev}, PKS 1749+096 \citep{Cui2021Resolving-the-i}, NGC 315 \citep{Park2021Jet-Collimation}, and PG 1553+113 \citep{Orienti2020Radio-VLBA-pola}. 
A triple-ridge structure composed of the limb-brightened components and an additional central bright component has been observed in the M87 jet \citep{Asada2016Indication-of-t, Hada2017Pilot-KaVA-moni, Walker2018The-Structure-a}.
The jet width profile along the distance from the black hole (BH) have been measured in several jets \citep{Asada2012The-Structure-o, Boccardi2016First-3-mm-VLBI, Boccardi2020Jet-collimation, Tseng2016Structural-Tran, Nakahara2019The-Cygnus-A-Je, Nakahara2020The-Two-sided-J, Hada2018Collimation-Acc, Nakamura2018Parabolic-Jets-, Kovalev2020A-transition-fr, Boccardi2021Jet-collimation, Park2021Jet-Collimation, Lu2023Nature, Cui2023Nature}.

It is crucial to reveal the formation mechanism of relativistic jets via studies of the jet launching region.
AGN jets are thought to be launched near a central supermassive BH.
The Event Horizon Telescope (EHT) has revealed the ring-like emission structure around the event horizon of M87 at 230 GHz \citep{Event-Horizon-Telescope-Collaboration2019First-M87-Eventa, Event-Horizon-Telescope-Collaboration2019First-M87-Eventb, Event-Horizon-Telescope-Collaboration2019First-M87-Eventc, Event-Horizon-Telescope-Collaboration2019First-M87-Eventd, Event-Horizon-Telescope-Collaboration2019First-M87-Evente, Event-Horizon-Telescope-Collaboration2019First-M87-Eventf}. 
However, an extended emission structure that connects to the jet seen in the lower-frequency observations has not been confirmed.
Future high-resolution observations are expected to reveal the extended images from the jet at its launching zone \citep{Doeleman2019Studying-Black-}.

The formation mechanism of the jets near a central supermassive BH is still under debate.
One of the plausible jet launching mechanism is the Blandford-Znajek process \citep[][]{Blandford1977Electromagnetic}.
The BH spin twists the magnetic field lines threading the BH, generate the Poynting flux extracting the rotation energy from the BH, and accelerate plasma to relativistic speed.
The dynamics of AGN jets have been studied both analytically and numerically.

Numerical general relativistic ideal magnetohydrodynamic (GRMHD) simulations have been performed to investigate the surroundings of a BH \citep[][]{Koide1998General-Relativ, De-Villiers2003Magnetically-Dr,  McKinney2004A-Measurement-o, Mizuno2006RAISHIN:-A-High, White2016An-Extension-of, Porth2019The-Event-Horiz}
\citep[see the review:][]{Mizuno2022GRMHD-Simulatio}.
Highly magnetized funnel regions, which are also indicated by analytical works on the observation data \citep{Kino2014Relativistic-El, Kino2015Magnetization-D}, are formed along the spin axis of the rotating BH, and the relativistically accelerated jets appear there.
However, such simulations have numerical difficulties in exactly calculating the highly magnetized jet regions, because of the significant contaminations due to artificial floors on the density and internal energy. 
Some simulations also have artificial ceilings on the magnetization parameter, the ratio of the electromagnetic energy flux to the rest-mass energy flux. 

Because of the problem mentioned above, a number of general relativistic radiative transfer (GRRT) calculations are carried out with neglecting the emission and absorption in the highly magnetized jet regions in GRMHD simulation data, in order to avoid artificial and uncertain appearance of the relativistic jets 
\citep[e.g.,][]{Dexter2009Millimeter-Flar,Dexter2012The-size-of-the,  Moscibrodzka2014Observational-a, Moscibrodzka2016General-relativ,
Gold2017Probing-the-Mag,
Davelaar2019Modeling-non-th,
Tsunetoe2020Polarization-im, Kawashima2023RAIKOU}.
In practice, 
\citet[][]{O-Riordan2018Observational-S} showed significant observational differences in the synthetic spectra, particularly in the optical and X-ray bands between the models in which the funnel is empty and the models where the funnel is filled with plasma, and \citet[][]{Chael2019Two-temperature} demonstrated synthetic images and spectra with different ceiling values of magnetization parameters.
These studies showed that the floor or ceiling values bring uncertainty on synthetic images, while this effect is not significant when the magnetization is not so much high in the highly magnetized region \citep{Chael2019Two-temperature, Event-Horizon-Telescope-Collaboration2021First-M87-Eventh}.

Alternative to the GRMHD simulation studies, analytical works has been carried out. While it is somewhat difficult to incorporate the time-dependent dynamics, it is possible to study the highly magnetized region of the relativistic jets. For example, force-free jets are studied \citep[e.g.,][]{Zakamska2008Hot-Self-Simila, Gracia2009Synthetic-Synch, Broderick2009Imaging-the-Bla} and improved models of 
\citet[][]{Takahashi2018Fast-spinning-B} and \citet[][]{Ogihara2019A-Mechanism-for} reproduced the observed jet structures in jets, e.g., limb-bright structure and triple ridge structure, respectively.

While the force-free approximation is appropriate for the extremely magnetized region, it is difficult to study the jets with their non-negligible internal and/or kinetic energy. Analytical GRMHD approaches have been, therefore, also developed to overcome this problem.
The basic equations are divided into parallel and perpendicular components to the magnetic field line. 
The parallel component is analytically solvable under the assumption of the magnetic flux function and four integral constants \citep{Bekenstein1978New-conservatio, Camenzind1986Centrifugally-d, Camenzind1986Hydromagnetic-f, Takahashi1990Magnetohydrodyn, Pu2015Steady-General-, Pu2020Properties-of-T, Ogihara2021Matter-Density-}.
The perpendicular component has analytic solutions in the force-free limit for the low BH spin case. 
They are called the monopole and parabolic Blandford \& Znajek perturbation solution \citep{Blandford1976Accretion-disc-, Blandford1977Electromagnetic}.
For higher BH spin cases,  \citet{Tanabe2008Extended-monopo} derived higher-order terms of the perturbation solutions, and they confirmed the analytically-derived total energy flux is consistent with the numerical results in the monopole field-line case.
\citet{Nathanail2014Black-Hole-Magn} conducted numerical force-free simulations, and they showed the field line configurations inside the fixed monopole and parabolic field lines.

In the steady and axisymmetric GRMHD formulation, \citet{Huang2019Toward-a-Full-M} and \citet{Huang2020Toward-a-Full-M} studied two-dimensional distributions of the electromagnetic field, the density, and velocity of plasma by iteratively solving the parallel component and the perpendicular component for the monopole and parabolic field line configuration. 
In the parabolic field line model, they introduced the "loading zone." The inner boundary of the loading zone is the null-charge surface, and the outer boundary is the surface where the outflow velocity obtained by solving the Bernoulli equation becomes zero. 
The outflow and inflow start at these surfaces. 
\citet[][]{Ogihara2021Matter-Density-} constructed the two-dimensional jet models without the loading zone by solving the Bernoulli equation analytically, and the trans-field force-balance numerically only at the separation surface of the inflow and the outflow. 
The force-balance is relatively well established near the separation surface and far from the BH.
It will be fruitful to demonstrate the appearance of the highly magnetized jets inside the jet funnels towards  future VLBI mission with high-resolution and high-sensitivity to detect the images of horizon-scale jets.

In this paper, we compute the synthetic radio images of highly magnetized relativistic jets inside the jets funnels based on the semi-analytic GRMHD models of \citet{Ogihara2021Matter-Density-}. 
We focus on the emission structure of jets in the vicinity of the BH horizon, where is thought as the origin of the jet.
As a first attempt, we demonstrate the observational features of highly magnetized jet based on the semi-analytic GRMHD model, rather than dedicating to tune the model parameters reproducing the observed image precisely.

This paper is organized as follows.
In \S \ref{sec:method}, we briefly introduce the GRMHD jet models and the GRRT code we use.
In \S \ref{sec:Main Results}, we show the resultant synthetic images and their characteristic features.
In \S \ref{sec:summary}, we discuss comparisons to the observed images and other studies.
Summary and prospects are presented in \S \ref{sec:summary}.

\section{Method} \label{sec:method}
To compute synthetic images of the semi-analytic GRMHD models of \citet{Ogihara2021Matter-Density-}, we perform the GRRT calculations  using \texttt{RAIKOU} code \citep{Kawashima2023RAIKOU}.
Here, we focus on the emission features of relativistic jets, treating the outside of the jets as a vacuum.

The spacetime geometry is given by the Kerr metric in the Boyer-Lindquist coordinates,
\begin{eqnarray}
    ds^2 = &-&\left(1-\frac{2Mr}{\Sigma}\right) c^2dt^2 - \frac{4aMr\sin^2\theta}{\Sigma} dt d\phi + \frac{\Sigma}{\Delta} dr^2 \nonumber \\ 
    &+& \Sigma d\theta^2 + \frac{[(r^2+a^2)^2-\Delta a^2\sin^2\theta]\sin^2\theta}{\Sigma} d\phi^2, \nonumber \\
    \label{eq:metric}
\end{eqnarray}
where $\Sigma=r^2+a^2 M^2 \cos^2\theta$,  $\Delta=r^2-2 M r+ a^2 M^2$.
$M$ is the BH mass.
The gravitational radius is $r_{\rm g} = GM/c^2$. 
$a$, $c$, $G$ are the dimensionless spin parameter, speed of light, gravitational constant. 
In Equation \ref{eq:metric}, we set $c=G=1$.
The radius of the event horizon is $r_{\rm H} = r_{\rm g} (1+ \sqrt{1-a^2})$.
The angular velocity of the BH is $\Omega_{\rm H} = a c/2r_{\rm H}$.

\subsection{GRMHD Jet Model} \label{subsec:GRMHD Jet Model}
\citet{Ogihara2021Matter-Density-} constructed steady, axisymmetric GRMHD jet models.
They assumed the fixed poloidal magnetic field shapes that mimic force-free analytic solutions and GRMHD simulation results. 
The field line shapes are defined by the flux function 
\begin{equation}
    \Psi(r,\theta) = C \left[ \left( \frac{r}{r_{\rm H}} \right)^\nu(1-\cos\theta) + \frac{1}{4} \epsilon \frac{r}{r_{\rm g}} \sin\theta \right], 
\end{equation}
where 
$C$ is the normalization factor which set the magnetic field strength is 10 G at the ergosphere of the outermost field line.
They only consider the field line threading the BH.
$\nu$ and $\epsilon$ are the model parameters controlling the poloidal field line shape.
When $\epsilon=0$, the field line configurations of $\nu=0$ and $\nu=1$ have the monopole and parabolic shape in the far zone, respectively.

There are four integral constants along the field line:
the total energy flux per particle $E$, the total angular momentum flux per particle $L$, the number density flux per magnetic flux $\eta$, and the so-called “angular velocity of the magnetic field” $\Omega_{\rm F}$ \citep[][]{Bekenstein1978New-conservatio}.
\citet{Ogihara2021Matter-Density-} determined the value of these integral constants by the four conditions: the regularity condition at the fast magnetosonic point in the outflow, the Znajek condition \citep[][]{Znajek1977Black-hole-elec}, the initial poloidal velocity at the separation surface, and the force-balance at the separation surface.
The separation surface is given by the surface where $\partial_p k_0 = 0$ for each field line, where $\partial_p$ is the directive differentiation along the field line, and $k_0 = -(g_{00} + 2\Omega_{\rm F} g_{03} + \Omega_{\rm F}^2 g_{33})$.
There are five free parameters in the models in \citet{Ogihara2021Matter-Density-}, which are the BH's dimensionless spin parameter $a$, the field line configurations $\nu$ and $\epsilon$, the poloidal four velocity at the separation surface, and the angular velocity of the outermost field line threading the BH.

In this paper, we use the jet models P1, P2, and P3 of \citet{Ogihara2021Matter-Density-}. 
The P1 model is the fiducial one, in which the BH spin parameter is $a=0.9$. 
The P2 and P3 model have $a=0.8,$ and $0.95$, respectively. 
$\nu=1$ and $\epsilon=10^{-4}$ for all the models.
The poloidal four velocity at the separation surface $(r=r_{\rm ss})$ is given by $u_p(r=r_{\rm ss}) = 10^{-3}$.
The angular velocity of the outermost field line threading the BH is given by $\Omega_F(\Psi=\Psi_{\rm max}) = 0.35\Omega_{\rm H}$, where $\Psi_{\rm max}$ is the value of the flux function at the outermost field line.
The force-balances between the field lines are numerically solved only at the separation surface from $\Psi=\Psi_{\rm max}$ toward the jet axis.
As a consequence, the distributions of the toroidal magnetic field, poloidal electric field, matter velocity, and density on each field lines are analytically obtained.
The other model parameters are the same among the models.
Notable differences between the models are below:
(i) The maximum value of magnetization parameter is higher in the model with a larger spin.
(ii) The density at the separation surface is higher in the jet edge and lower near the jet axis when the spin is larger \citep[see Figure 4 in][]{Ogihara2021Matter-Density-}.
We use the data of the GRMHD models from the radius of the ergosphere to $\sim 10^5 r_{\rm g}$.
We do not consider the emission from the outside of the jet boundary $\Psi = \Psi_{\rm max}$ and inside the ergosphere.
Near the jet axis, in the white region in Figure \ref{fig:2d}, the electromagnetic field, velocity field, and density are filled with the values from the closest cell.

\subsection{GRRT Calculation} \label{subsec:GRRT calculation}
The nonthermal synchrotron radio images of the GRMHD jet models are computed by using a general relativistic multiwavelength radiative transfer code \texttt{RAIKOU} \citep{Kawashima2023RAIKOU}. 
In \texttt{RAIKOU} code, two types of the algorithms are implemented: (i) the observer-to-emitter ray-tracing algorithm for the efficient calculations of the non-Compton-scattered photons, and (ii) the emitter-to-observer Monte-Carlo algorithms for the broadband multiwavelength calculations with including Compton processes.
In this work, we employ the observer-to-emitter ray-tracing algorithm for the efficient calculations of the synchrotron radio images of the jets.

In this paper, we use the parameter values dedicated to M87.
We set the BH mass and the distance to the BH to be $6.5\times 10^9 M_\odot$ and 16.9 Mpc \citep[][]{Event-Horizon-Telescope-Collaboration2019First-M87-Eventf}, respectively.
In our fiducial model, the viewing angle from the jet axis is $\theta_{\rm view} = 15^{\circ}$.
In addition, we also examine $\theta_{\rm view} = 5^{\circ}, 45^{\circ}$, and $85^{\circ}$ to generally understand the observational signature of our jet model via the comparison among them.

We assume that all the particle consists of single power-law non-thermal electrons and consider the synchrotron emission and absorption from them.
We mainly study the models with the steeper power-law index $p = -3.5$ \citep[e.g.,][]{Dexter2012The-size-of-the, Kawashima2021}, where $N(\Gamma_{\rm e}) \propto \Gamma_{\rm e}^p$. $N$ is the particle number density in the phase space and $\Gamma_{\rm e}$ is the electron's Lorentz factor. 
We also discuss the dependence of the jet appearance on the power-law index through the comparison with the model with the shallower power-law index $p=-1.1$.
For all models, the minimum and maximum values of $\Gamma_{\rm e}$ are $30$ and $10^6$, respectively, which are the same as \cite{Kawashima2021}.

In our GRMHD jet model, the magnetic field strength is set in such a way that the field strength on the outer boundary of the ergosphere on the equatorial plane is $B=10$ G as is described in \S \ref{subsec:GRMHD Jet Model}.
The non-thermal electron density is converted to c.g.s unit by multiplying the square of the normalization factor of the magnetic field strength.

\section{Results} \label{sec:Main Results}

We first present the results of the GRRT calculation for the fiducial model, in which $a=0.9$, $\theta_{\rm view}=15^\circ$, $p=-3.5$, and the observational frequency is 230 GHz.
The emission sources that contribute to the observed images
are shown.
Then, we also show the dependence of synthetic images on the viewing angle, observational frequency, BH spin, and the power-law index of the electrons.
Again, it would be worth mentioning here that the scope of this paper is demonstrating the appearance of the radio image of the highly magnetized, semi-analytic GRMHD jet model of \citet{Ogihara2021Matter-Density-}. 
The detailed model parameters are not tuned for reproducing the observed radiative fluxes at $\sim 0.5$ Jy both at 230 GHz by EHT \citep{Event-Horizon-Telescope-Collaboration2019First-M87-Eventa} and at 86GHz by GMVA, ALMA and GLT \citep{Lu2023Nature} in M87, but the magnetic field strength of the outermost field line of the jet at the equatorial plane are fixed to 10 G as  mentioned in \S 2. 
In our fiducial model, the resulting radiative fluxes are $\sim 2$ and $\sim 1.8$ Jy at 230 and 86 GHz, respectively.

\subsection{Synthetic Images} \label{subsec:Synthetic Images}

\begin{figure*}

    \centering

    \includegraphics[width=\linewidth]{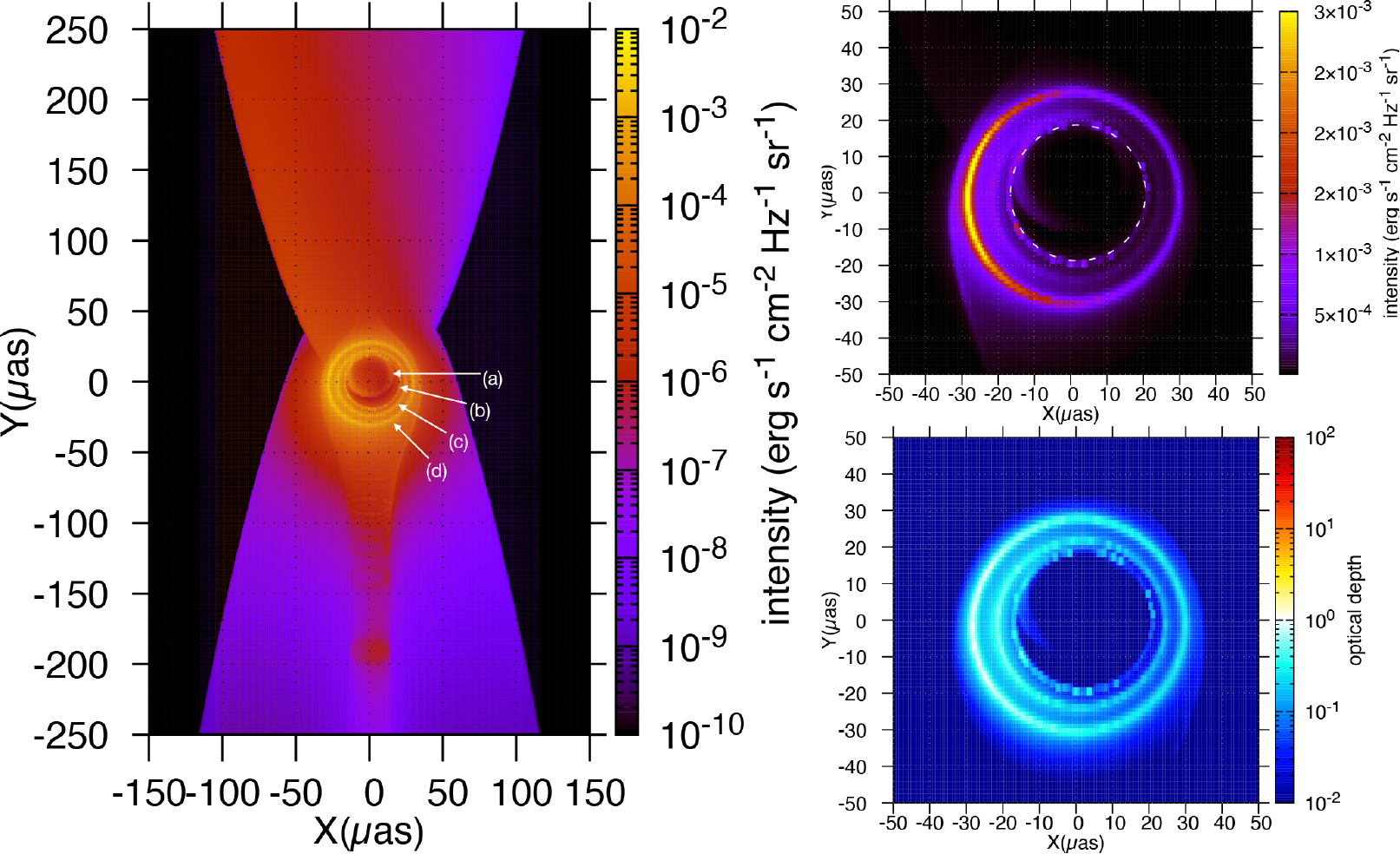}
    
    \caption{
    The synthetic images in logarithmic (left panel) and linear scales (right upper panel) with the fiducial parameter set, i.e., BH spin parameter $a=0.9$, viewing angle $\theta_{\rm view} = 15^{\circ}$, power-law index $p=-3.5$, and the observational frequency 230 GHz. The optical depth map is plotted in the right lower panel.
    The white dashed circle in the right upper panel represents the position of the photon ring. 
    The several thin, ring-like red discontinuities seen in the tear-drop shape in the counter jet between around $Y\sim -100$ and $-200 \mu$as are the artifacts resulting from the limited resolution of the semi-analytic model used to generate this image and have no physical meaning.
    }
    \label{fig:fiducial-model-image}
\end{figure*}

Figure \ref{fig:fiducial-model-image} shows the intensity and the optical depth maps of the fiducial model. 
In each panel, the zero-angular-momentum photons pass through the origin of the observer screen $(X,Y)=(0,0)$, i.e.,the center of the BH is projected  to the origin of the screen.
The projected length of $r_{\rm g}$  is $\arctan(r_{\rm g}/D) \sim 3.8 \mu$as on the screen. The de-projected length of  $r_{\rm g}$ in the direction of the BH spin axis, i.e., the jet-direction, corresponds to $\arctan(r_{\rm g}\sin(\theta_{\rm view})/D) \sim 0.98 \mu$as in the $Y$ direction on the screen, when the viewing angle is $\theta_{\rm view} = 15^\circ$.
The log-scale intensity map (left panel) is composed of the bright four rings, a bright teardrop shaped component extending downward from the bright rings, and the broadly extended component that outlines the entire jet shape. 

We label four rings 
as the ring (a), (b), (c), and (d) from the smallest to the largest.
The rings (a) and (b) are due to radiation from the bottom of the separation surface in the approaching jet. 
Meanwhile, emission from the bottom of the separation surface in the counter jet produces rings (c) and (d).
The emission from near the separation surface contributes to the formation of the teardrop shaped bright region in the counter jet.
We will discuss the details later.

The relativistic beaming is the cause of asymmetry in the broadly extended regions.
The approaching jet is brighter than the counter jet because approaching jet (counter jet) is moving toward (away from) the observer, except in the inflow regions on the BH side of the separation surface.
The asymmetry of the broadly extended regions with respect to the $Y$-axis is due to the relativistic beaming effect by the toroidal component of the fluid velocity. 

The upper right panel of Figure \ref{fig:fiducial-model-image} is the linear-scale intensity map cropped by $100 \mu {\rm as} \times 100 \mu {\rm as}$ from the log-scale intensity map. 
The white dashed circle indicates the position of the photon ring.
Of the two clearly visible rings, 
the inner one is ring (c) and the outer one is ring (d).
The rings (a) and (b) are too dark to see clearly in this panel.
The diameter of the outermost brightest ring (d) is $\sim 60 ~\mu{\rm as}$, which is $\sim 1.5$ times larger than that of the ring-like image observed by EHT and similar with that observed by GMVA \citep{Lu2023Nature}.
The possible application of our model and consistency with the observations will be discussed in {\S}\ref{sec:86GHz}  and {\S}\ref{sec:summary}

In the lower right panel of Figure \ref{fig:fiducial-model-image}, 
the optical depth map is demonstrated.
The optical depth becomes nearly unity at the bright rings, although the other area is optically thin.
This indicates that the  ray forming the bright rings pass through the high opacity (i.e., high emissivity) regions, with the larger path length due to the light bending effects.

\begin{figure}
    \centering
    
    \includegraphics[width=\linewidth]{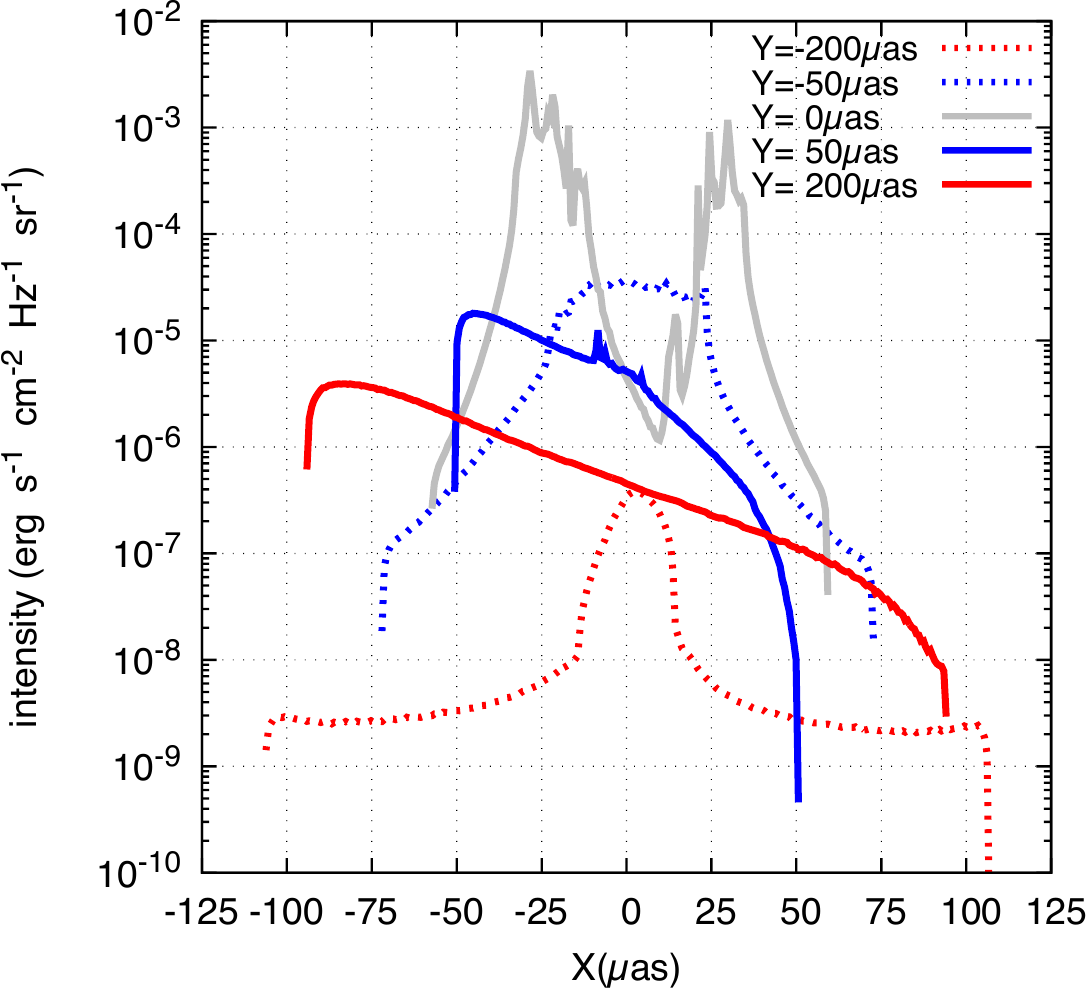}

    \caption{The intensity slices at $Y=-200,$ $-50,$ $0,$ $50,$ and $200 \mu as$ of Figure \ref{fig:fiducial-model-image}.}
    \label{fig:intensity-slice}
\end{figure}

In order to quantitatively understand the distribution of the intensity, we show the intensity slices at $Y=-200,$ $-50,$ $0,$ $50,$ and $200 \mu {\rm as}$ in Figure \ref{fig:intensity-slice}. 
The positions of two intensity peaks, $X \sim \pm 25\mu{\rm as}$ for the slice at $Y=0$ (gray line)
corresponds to the four rings.
In the slice at $Y=-50\mu {\rm as}$ (blue dotted) and $Y=-200 \mu {\rm as}$ (red dotted), 
the bright parts in $-25\mu {\rm as} \lesssim X \lesssim 25\mu {\rm as}$ 
and in $-15\mu {\rm as} \lesssim X \lesssim 15\mu {\rm as}$ 
are the teardrop-shaped component.
The relatively less luminous part outside of them corresponds to the extended emission region
in the counter jet.
In the $Y=50\mu {\rm as}$ slice and in the $Y=200\mu {\rm as}$ slice,
most of the radiation comes from the outflowing matter in the approaching jet.
The asymmetry with respect to the $Y$-axis is due to the relativistic beaming effect as we have mentioned above.
The relativistic beaming effect also makes the intensity higher in the approaching jet ($Y>0$) than in the counter jet ($Y<0$). 
In the distant region from the BH (e.g., $Y=200$ $\mu {\rm as}$, this effect is more efficient due to the acceleration of the relativistic jets.
The gravitational lensing also affects the images.
The apparent jet width of the counter jet is wider than that of the approaching jet, i.e., one may find that the brighter emission regions are wider in the slices of $Y=-50 ~\mu {\rm as}$ and $-200 ~\mu {\rm as}$ than those of  $Y=50 ~\mu {\rm as}$ and $-200 \mu {\rm as}$, respectively.
It should be noted that the small spikes at around $-15 ~\mu {\rm as} \lessapprox X \lessapprox 7.5~ \mu {\rm as}$ in the $Y=50 ~\mu {\rm as}$ slice represent the contamination by the emission from the counter jet.

\begin{figure}
    \centering

    \includegraphics[height=200mm]{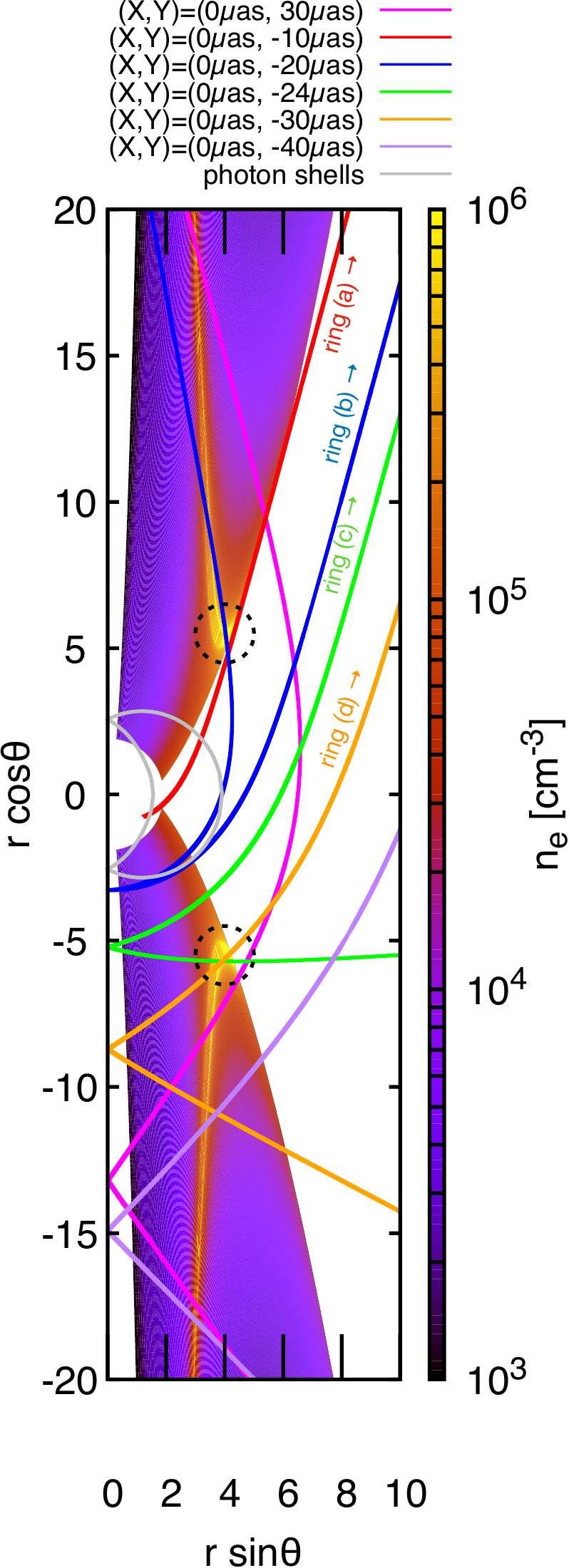}
    
    \caption{
    The two-dimensional color map of $n_e$.
    The trajectories of the ray which reaches the ring (a),(b),(c),(d) and $(X,Y) = (0\mu {\rm as}, -10\mu {\rm as})$, $(0\mu {\rm as}, -20\mu {\rm as})$, $(0\mu {\rm as}, -24\mu {\rm as})$ and $(0\mu {\rm as}, -30\mu {\rm as})$ on the observer screen are overlaid. 
    We also show the other two trajectories that reach $(X,Y)=(0\mu {\rm as}, 30\mu {\rm as})$ and $(X,Y)=(0\mu {\rm as}, -40\mu {\rm as})$ on the observer screen for reference.
    The black lines represent the separation surface. The gray lines are the photon shells \citep{Teo2003}, and the black dotted circles point the area of the bottom of the separation surface.
    The U-shape patterns near the jet axis are the printing artifacts.
    }
    \label{fig:2d}
\end{figure}

Here we show in detail the origin of 
the bright four rings and a bright teardrop shaped 
component.
Figure \ref{fig:2d} displays the distribution of $n_e$ and some trajectories of the ray.
The trajectory for ring (a) (red line) leads from the observer screen to the BH. The photons emitted at the bottom of the separation surface 
(dotted circle) 
immediately goes out of the jet region and reaches the observer screen.
On the other hand, the blue line indicates that photons emitted at the bottom of the separation surface pass through the region of the counter jet before reaching the screen.
This is due to the gravitational lensing.
As a result, a slightly larger ring than ring (a) appears on the screen.
Thus, the bottom of the separation surface in the approaching jet, which is a single ring shape, creates two rings on the observer screen.
We note that the image of the ring (b) is almost identical to the photon ring (see Figure \ref{fig:fiducial-model-image}), because the ray passes through the region inside the photon shell (gray lines in Figure \ref{fig:2d}) , where the unstable circular orbit forming the photon ring exists.
The bottom of the separation surface in the counter jet also produces two rings.
The trajectory indicated by the orange line reach the observer screen directly from the bottom of the separation surface. This corresponds to the ring (d).
The ring (c) is generated by photons, which emitted at the bottom of the separation surface, pass through the interior of the counter jet and reach the observer screen.

We also find that the trajectory indicated by the purple line clearly passes through the separation surface in the counter jet.
Because of the large emissivity at the separation surface, the bright teardrop shaped component appear at the lower part of the screen ($Y<0$).
Light rays passing through the separation surface also reach the upper part of the screen ($Y>0$). However, this emission is weaker than the boosted component of the approaching jet, so the teardrop shaped component does not appear.
We note that the asymmetry of the teardrop-shaped
bright region with respect to the $Y$-axis 
(shown in Figure \ref{fig:fiducial-model-image})
is due to the relativistic beaming effect 
by the toroidal component of the fluid velocity.

\subsection{Decomposing the emission structure}

\begin{figure*}
    \centering 
    
    \includegraphics[width=\linewidth]{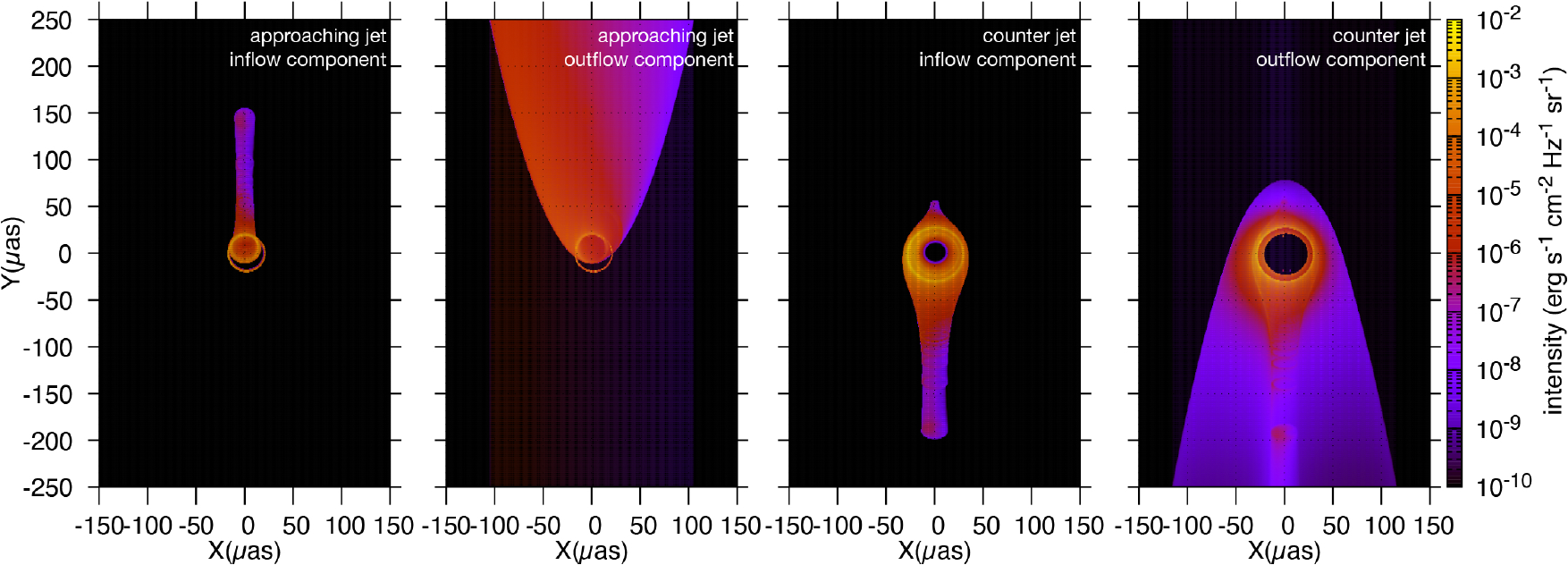}

    \caption{
    The intensity maps of the emission from the inflow component of the approaching jet, the outflow component of the approaching jet, the inflow component of the counter jet, and the outflow component of the counter jet from the left to the right panel. 
    These are the decomposed intensity map of the left panel of Figure \ref{fig:fiducial-model-image}.
    As in Fig. 1, the several thin, ring-like red discontinuities around $Y\sim-100$ and $-200\mu$as are the artifacts.
    }
    \label{fig:inflow-outflow-approaching-counter}    
\end{figure*}

To decompose the emission structure of Figure \ref{fig:fiducial-model-image},
we show the intensity maps of the emission  originated from the four components: (i) inflow component of the approaching jet, (ii) the outflow component of the approaching jet, (iii) the inflow component of the counter jet, and (iv) the outflow component of the counter jet in Figure \ref{fig:inflow-outflow-approaching-counter}.

The extended bright region in the second image from the left
is originated from the outflow component in the approaching jet,
which extends outward starting from the separation surface.
On the other hand, the vertical structure in the leftmost image 
appears as a consequence of the emission from dense gas near the separation surface.
The intensity in this vertical region is smaller than that 
in the extended region of the second image from the right and therefore 
does not appear in the approaching jet in Figure \ref{fig:fiducial-model-image}.
Two bright rings in the two left-hand images
are rings (a) and (b).
The density (emissivity) in the region near 
the bottom of the separation surface is so large 
that emission from this region forms bright rings.
Since the fluid velocity in this region is 
positive on one side and negative on the other across the surface, 
the rings appear in both of the two left-hand panels.
The extended emission region 
as well as the  vertical region
connects to the ring (a).
This means that the ring (a) is the direct emission image of the separation surface.
Conversely, the ring (b) does not smoothly connect to other luminous regions 
because it is formed  by photons that pass near the unstable orbit, which exists inside the photon shell (gray lines depicted in Figure \ref{fig:2d}), and the indirect image is formed.

Two rings (c) and (d)
appear in the 
two right-hand images.
They are formed by radiation from the bottom of the separation surface
in the counter jet.
The extended region in the rightmost image 
is  produced by the radiation from the outflow component in the counter jet,
and the emission from the separation surface in the counter jet
is responsible for the teardrop-shaped region in the second image from the right. 
The extended emission region is darkened by the  relativistic beaming effect, 
so the teardrop-shaped region is visible in Figure \ref{fig:fiducial-model-image}.
We note in the rightmost image
that the extended region largely 
extends into the $Y>0$ regime 
due to the gravitational lensing.

\subsection{Parameter Dependencies}
In this subsection, we change 
the viewing angle, observation frequency, BH spin, and energy distribution of electrons, and 
compare the difference between the synthetic images.

\subsubsection{Viewing angle dependence}

\begin{figure*}
    \centering 
    
    \includegraphics[width=\linewidth]{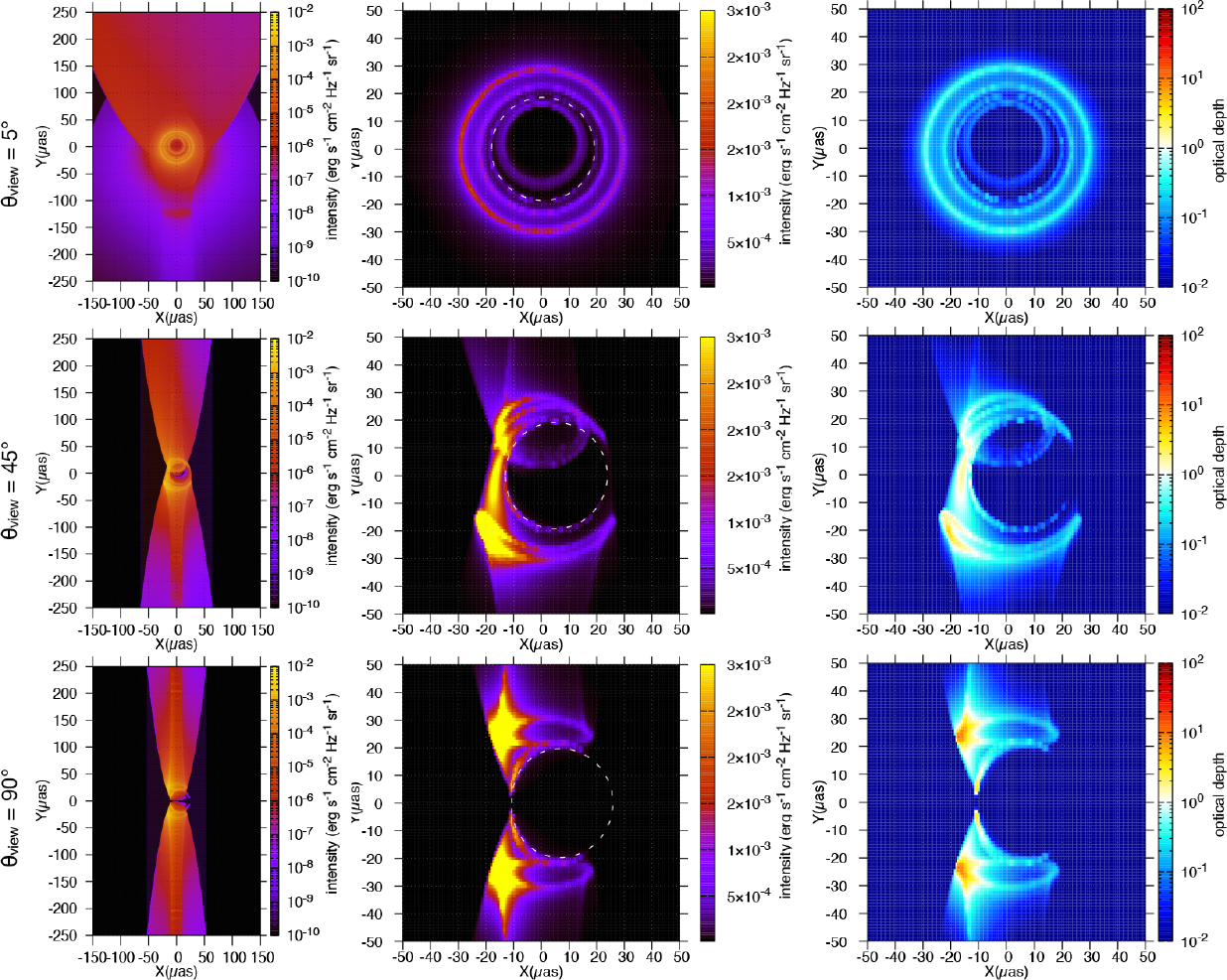}

    \caption{The intensity maps and the optical depth maps when the viewing angle $\theta_{\rm view}$ is $5^\circ$ (top-row panels), $45^\circ$(middle-row panels), and $90^\circ$ (bottom-row panels). The left-column panels are the log-scale intensity maps, the middle-column panels are the linear-scale intensity maps, and the right-column panels are the optical depth maps. The white dotted circles in the linear-scale intensity map represent the location of the photon ring.}
    \label{fig:viewing-angle-dependence}
    
\end{figure*}

Figure \ref{fig:viewing-angle-dependence} shows the intensity maps and the optical depth maps when the viewing angle $\theta_{\rm view}$ is $5^\circ, 45^\circ,$ and $90^\circ$.
As the viewing angle becomes larger, 
the broadly extended luminous region,
that outlines the entire jet shape becomes, narrower.
When $\theta_{\rm view}=5^\circ$, the asymmetry with respect to the $Y$-axis is less significant compared to the other viewing angles.
This is because the relativistic beaming via the toroidal velocity does not work so well.
Conversely, when $\theta_{\rm view}=90^\circ$,
the relativistic beaming is most effective
and the left side is brighter.
The shape of the ring is closest to circular when 
$\theta_{\rm view}=5^\circ$, 
and the shape is distorted as 
$\theta_{\rm view}$ approaches $90^\circ$ because of the frame-dragging effect.
The peak intensity and 
the peak optical depth also increase 
with  $\theta_{\rm view}$.
For the case of $\theta_{\rm view}=45^\circ$ 
and $90^\circ$,
the peak position of the 
intensity and the optical depth 
corresponds to the bottom of the separation surface,
where the toroidal component of the velocity is dominant, and the relativistic beaming effect enhances the intensity of the
left side.

\subsubsection{Frequency dependence} \label{sec:86GHz}

\begin{figure*}
    \centering     
    \includegraphics[width=\linewidth]{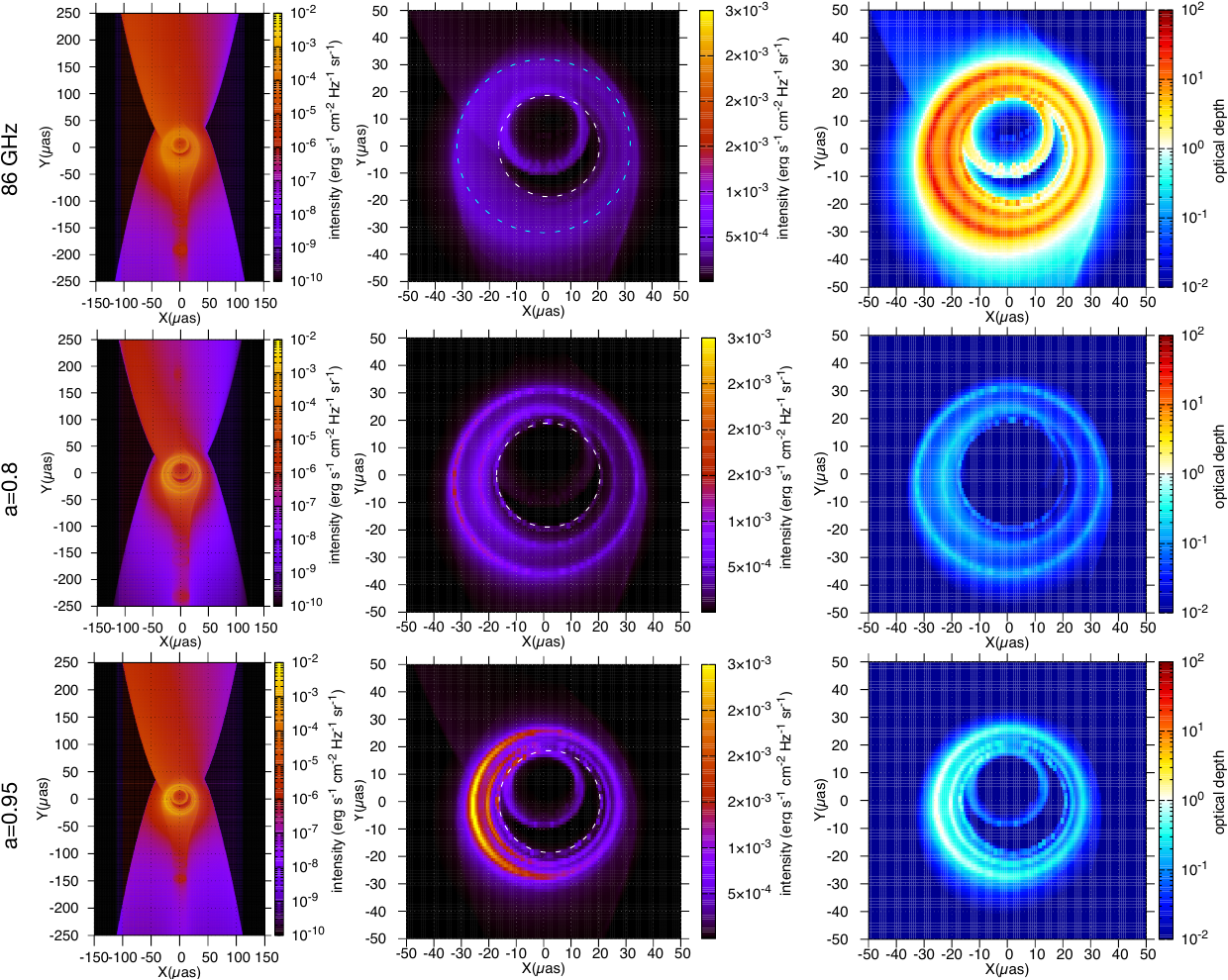}
    
    \caption{The intensity maps and the optical depth maps when the observational frequency is 86 GHz (top-row panels), and when the BH spin parameter $a$ is 0.8 (middle-row panels) and 0.95 (bottom-row panels). The left-column panels are the log-scale intensity maps, the middle-column panels are the linear-scale intensity maps, and the right-column panels are the optical depth maps. The white dotted circles in the linear-scale intensity map represent the position of the photon ring. The diameter of the cyan dotted circle in the top-middle panel is 64$\mu$as, which is almost as large as the ring-like images observed with GMVA, ALMA, and GLT at 86 GHz \citep{Lu2023Nature}.}
    \label{fig:86ghz-spin-dependence}
    
\end{figure*}
The intensity maps and optical depth map when the observational frequency is 86 GHz are shown in the top-row panels of Figure \ref{fig:86ghz-spin-dependence}.
Since the absorption opacity for the synchrotron 
increases with the decrease of the frequency,
the optical depth is drastically enhanced
and highly exceeds the unity,
especially in and around the photon ring and the ring images of the separation surfaces
(see the top-right panel).
In the top-middle panel,
the rings are relatively faint due to absorption by the optically thick medium against the synchrotron  process.
In the top-left panel, we can see that 
the bright rings are blurred compared to 
the image at 230 GHz.
The central bright area 
would correspond to the core of M87.
The size of this bright region (the diameter $\sim 60 ~\mu {\rm as}$)
is roughly consistent with the recently observed 86 GHz ring image \citep{Lu2023Nature}, whose diameter is $\sim 64 \mu{\rm as}$.

The caveat is that the diameter of the ring-like image at 230 GHz is the same as that at 86 GHz in our model, while the EHT has not detected $60 \mu{\rm as}$ ring-like image.
However, as roughly estimated below,
this is expected to be solved if we construct the GRMHD jet model with optical depth being thinner against synchrotron absorption (e.g., lower density GRMHD jet). 

If the plasma is optically thin, 
the ratio of the radiative flux at 230 GHz to 86 GHz originated from the outermost brightest ring will be $F_{\nu}(\nu = {\rm 230 GHz})/F_{\nu} ({\rm 86 GHz}) = ({\rm 230 GHz}/{\rm 86 GHz})^{(1-p)/2} \sim 0.29$.
If $F_{\nu} ({\rm 86 GHz}) \sim 0.5 ~{\rm Jy}$ as being observed by GMVA, ALMA and GLT \citep{Lu2023Nature}, 
the radiative flux at 230 GHz is $F_{\nu}({\rm 230 GHz}) \sim 0.15 {\rm Jy}$, which is $\sim 30\%$ of that observed by EHT \citep{Event-Horizon-Telescope-Collaboration2019First-M87-Eventa}.
Since the diameter of our ring ($\sim 60 ~\mu {\rm as}$) is roughly 1.5 times larger than the EHT ring, the typical intensity of the outermost ring will be $\sim (1.0/1.5^2) \times 30\% \sim 10\%$, which is marginally lower than the detection limit of EHT due to its dynamical range.
Therefore, the outermost ring can disappear in the current ability of the EHT consistently, if the plasma is set to be optically thin against the synchrotron absorption. 
This will be confirmed in the future work. We note that the change of the model parameters of GRMHD jet does not prominently affect the diameter of our  outermost brightest ring-like images, since they are mainly regulated by the black hole spin.

\subsubsection{BH spin dependence}

The middle-row and bottom-row panels of Figure \ref{fig:86ghz-spin-dependence} are the intensity and the optical depth maps 
of the models with the other BH spin parameter $a = 0.8$ and $0.95$, respectively.
The GRMHD models, P2 ($a=0.8)$ and P3 ($a=0.95$), 
in \citet[][]{Ogihara2021Matter-Density-}
is here employed.
The separation surface and the unstable photon orbit 
become closer to the BH
as the spin parameter increases, 
so that the size of the bright rings decreases.
This ring-size feature is  consistent with the previous work using a toy model of the separation surfaces \citep{Kawashima2021}.
The peak intensity and the optical depth is higher 
when the spin value is larger.
Hence, small and clear rings appear 
for the case of $a=0.95$.
The structure of the extended emission region 
is not so sensitive to the spin parameter
(see left images).

\subsubsection{Dependence on the energy distribution of the electrons}

\begin{figure*}
    \centering 
    
    \includegraphics[width=\linewidth]{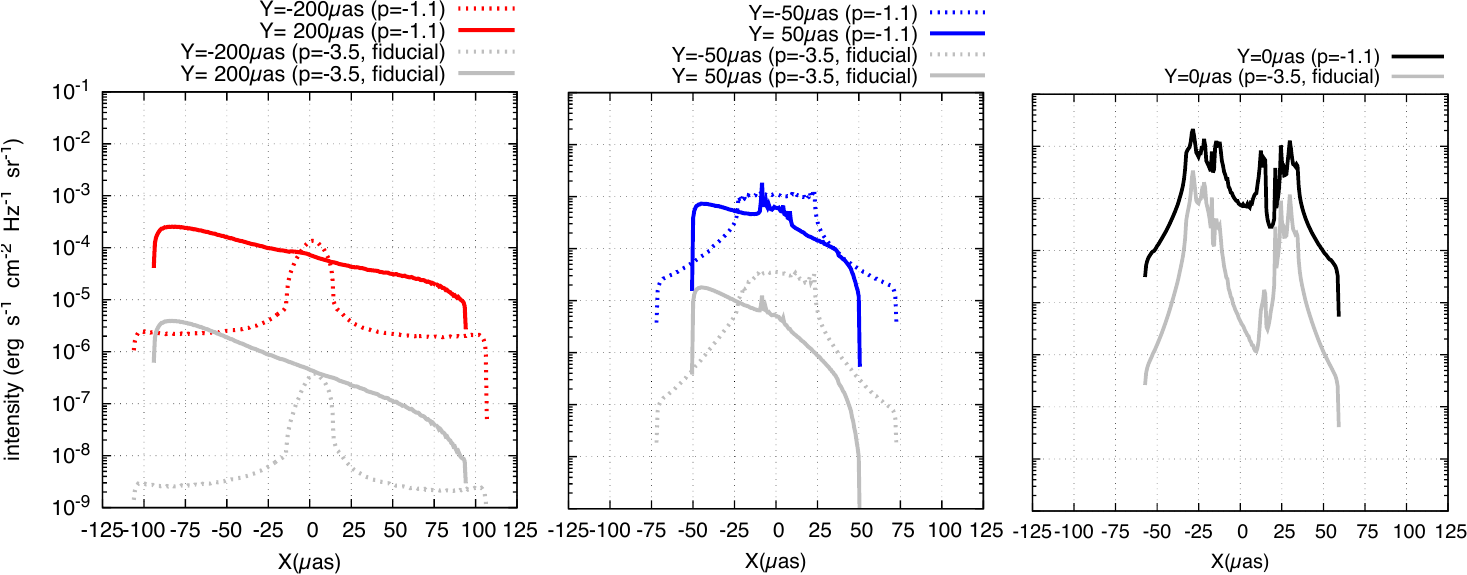}

    \caption{The intensity slices at $\pm 200\mu {\rm as}$ (left panel), $\pm 50 \mu{\rm as}$ (middle panel), and $0\mu{\rm as}$ (right panel) of the image at 230 GHz when the power-law index of the electron energy distribution is $p = -1.1$ and $-3.5$. The gray lines are the slices of the $p=-3.5$ (fiducial) model, and the colored lines are the $p=-1.1$ model.}

    \label{fig:power-law-dependence}
    
\end{figure*}

Figure \ref{fig:power-law-dependence} shows the intensity slices when the power-law index of the electrons' energy distribution is $p = -1.1$ and $-3.5$ (fiducial). 
The intensity of the $p=-1.1$ model is larger than the $p=-3.5$ model in all the slices since the large number of high-energy photons present in the model of $p=-1.1$.
The intensity contrast in the image becomes smaller when the energy distribution gets harder, 
because the observable intensity of the synchrotron photons emitted by  nonthermal electrons with the power law index $p$ is $I_{\nu} \propto \delta^{(5-p)/2}$, where $\delta = \nu/\nu^{\prime}$ is the relativistic beaming factor, $\nu$ is the photon frequency measured by observer, $\nu^{\prime}$ is the photon frequency measured in the fluid-rest frame, and $I_\nu$ is the specific intensity measured in the observer frame.
The dependence of the observed intensity on $\delta$ is weaker when the power-law index is harder.
In other words, the number density of the synchrotron-emitting electrons at the observing photon frequency becomes less different between the boosted area and the deboosted area, when the energy distribution gets harder.
The slope of the $Y=200\mu{\rm as}$ slice is shallower in the $p=-1.1$ model than the $p=-3.5$ model as well as the $Y=50\mu{\rm as}$ slice, so that the above discussion can be confirmed. 

Above, we mentioned the appearance of the weakly asymmetric intensity map with respect to $X=0$ in the harder electron distribution model, the same discussion can be applicable to understand the intensity asymmetry with respect to $Y=0$ (i.e., the ratio of approaching jet to the counter jet)
The intensity ratio between the $Y=-200 ~ \mu{\rm as}$ slice and the $Y=200 ~ \mu{\rm as}$ slice, and between $Y=-50 ~ \mu{\rm as}$ slice and the $Y=50\mu{\rm as}$ slice is also smaller.
For the $Y=0$ slices, it should be noted that the ring components suffer the absorption efficiently, and are in the optically thick regime when $p=-1.1$, which also contributes the shallower contrast image.

\section{Conclusions \& Discussion} \label{sec:summary}

We demonstrated the synthetic radio images inside highly magnetized jet funnel based on the steady, axisymmetric, and semi-analytic GRMHD models of \citet{Ogihara2021Matter-Density-}, where the plasma accelerate to the relativistic speed from the separation surface of the inflow and outflow.
Here the GRRT calculations, which take into account the synchrotron emission/absorption via the single power-law non-thermal electrons, are performed using the \texttt{RAIKOU} code \citep{Kawashima2023RAIKOU}, assuming a vacuum outside the jets.

For the case of $a=0.9$, $\theta_{\rm view}=15^\circ$, $p=-3.5$, the synthetic image at 230 GHz consists of 
the bright four rings, a bright teardrop shaped structure
extending downward from the rings, and the broadly extended components that outlines the entire jet. 
The four rings are formed by the emission of the 
bottom of the separation surface, which is the 
strongest radiation source in our semi-analytic GRMHD models.
The plasma stagnates on and around the separation surfaces and the density of the plasma is high at the bottom of the separation surface near the jet funnel \citep[see Figure 3 in][]{Ogihara2021Matter-Density-}.
The totally four rings are formed via the emission from the bottom of the separation surfaces in the approaching and counter jet regions.
The ring with the smallest diameter is  the direct image of the emission from the bottom of the separation surface in the approaching jet.
The ring with the second from the smallest is formed by the photons passing through the region near the unstable circular orbit around the BH, i.e., the trajectory is strongly bent by the gravity of the BH, after the emission from the bottom of the separation surface in the approaching jet side.
The other two rings are the gravitationally lensed images of the mission from the bottom of the separation surface in the counter jet region.

The diameter of the brightest, outermost ring is approximately 1.5 times that of the photon ring diameter at 230 GHz.
We discussed in \S\ref{sec:86GHz} that this can be the origin of the ring-like feature observed by GMVA, ALMA, and GLT at 86 GHz \citep{Lu2023Nature}, whose diameter is also $\sim$ 1.5 times  larger than that of EHT ring image. 
Here, we discuss another possibility. It may also indicate
that the BH mass is overestimated by about 1.5 times if the EHT captures this ring rather than the photon ring. In this case, the photon ring is hidden in the BH shadow.
On the other hand, our result that the sizes of the bright rings at both 86 GHz and 230 GHz are nearly identical is inconsistent with the observations. 
It has been reported by the observations with GMVA, ALMA, and GLT that the ring diameter at 86 GHz is approximately $64^{+4}_{-8} \mu\mathrm{as}$, which is about 1.5 times larger than the observed ring at 230 GHz. This contradiction may be caused by that the disk radiation is ignored in the present work. If disk radiation predominates over separation surface radiation at 86 GHz, it could result in the appearance of a larger ring in the 86 GHz image than what is observed in this study. 
Here we note that due to the frequency dependence of the synchrotron opacity, the disk is likely to be brighter at 86 GHz than at 230 GHz.

Radiation from the separating surface in the counter jet 
other than the bottom forms the teardrop-shaped structure.
The outflow components in the approaching and counter jets are responsible for the broadly extended components.
The relativistic beaming effect due to the toroidal velocity induces asymmetry about the $Y$-axis, and that due to the poloidal velocity produces asymmetry with respect to the $X$-axis.

The apparent jet width is narrower as 
$\theta_{\rm view}$ approaches $90^{\circ}$.
Then, the shape of the bright rings distorted
and part of the bottom of the separation surface is luminous due to the relativistic beaming effect and the frame-dragging effect.
It was also found that the ring signature becomes less distinct because of the effective absorption
when the observed wavelength is 86 GHz.
We find that the rings tend to become smaller and clearer 
as the spin parameter increases.
In addition, 
we found that, when $p=-1.1$ is employed, the entire image becomes brighter and the difference between brightness and darkness becomes smaller.

Although we, in this study, only focus on the emission structure in $Y< 250 ~\mu{\rm as}$,
the observed limb-brightened features 
in $Y \gtrsim (1-10) ~{\rm mas}$ region
would be explained by the force-free models 
(\citet{Ogihara2019A-Mechanism-for}
see also \citet{Takahashi2018Fast-spinning-B}).
In the force-free models, a drift velocity is used as the fluid velocity,
and the Lorentz factor increases linearly with the cylindrical radius. 
As the poloidal velocity increases, the toroidal component of the velocity decreases, which decreases the asymmetry of the jet image.
In addition, the magnetic field and gas density are 
larger near the jet edge than around the axis.
Thus, the limb-brightened structure is expected to appear.

In this study, we assumed the all the particles are non-thermal electrons, and used the single power-law distribution for the electron energy spectrum in the entire jet region.
However, the spectrum should be determined by particle acceleration and cooling, 
as well as by electron-positron pair creation.
The pair creation by ambient photons originated from surrounding disk or corona can produces $\sim$ MeV pairs \citep{Kimura2020Hadronic-High-e}.
The distribution of the pairs have been studied by \citet{Moscibrodzka2011Pair-Production} and  \citet{Wong2020Pair-Drizzle-ar}.
Combination with the inverse-Compton scattering may create higher energy particles.
The unscreened electric field parallel to the magnetic field accelerate plasma particles (so-called "gap acceleration"). 
The gap appears at the null-charge surface rather than the separation surface according to GR particle-in-cell (PIC) simulations \citep{Levinson2018Particle-in-cel, Kisaka2020Comprehensive-A, Kisaka2022The-Response-of, Parfrey2019First-Principle, Crinquand2020Multidimensiona, Crinquand2021Synthetic-gamma}.
Recently, it is suggested that the injection of the charged particles inside the jet funnel can be realized via the cascades triggered by magnetic reconnection near the equatorial plane of the magnetically arrested disk in the vicinity of the BH \citep[][]{Kimura2022, Hakobyan2023}, motivated by the recent GRMHD simulations with remarkably high spatial resolution \citep{Ripperda2022}.
The change of synthetic images depending on the difference of the electron energy distribution should be studied in future studies.

In this study, the density is high at the edge of the separation surface 
\citep[as it is also shown in Figure 3 in][]{Ogihara2021Matter-Density-},
and the emission at the point contributes most for the brightness of the four rings. 
Therefore, the emission would weaken if the turbulence in the disk or wind occurs and the density at the edge decreases, although the detailed influence of the turbulence should be dealt with in another study.

The jet model used in this study is axisymmetric and does not include time variation. If a non-axisymmetric structure is generated, the resulting images could be modified. 
In practice, time variabilities of GeV-TeV gamma-rays are exhibited in M87 with timescale of a day \citep[e.g.,][]{Acciari2008}  in IC310 with several minutes \citep[e.g., ][]{Aleksic2014Sci}.
Incorporating time variability and non-axisymmetric structure into the model as well as including the emission and absorption via the accretion flow to generate a more realistic images is left as important future works.

\begin{acknowledgements}
We thank K. Toma for fruitful discussion. The numerical computations in this work were carried out at the Yukawa Institute Computer Facility. This work was supported by JSPS KAKENHI grant Nos. JP18K13594, JP23K03448, JP23H00117 (T.K.), JP18K03710, 21H01132, and JP21H04488 (K.O.). This work was also supported by “Program for Promoting Researches on the Supercomputer Fugaku” (Structure and Evolution of the Universe Unraveled by Fusion of Simulation and AI; Grant Numbergrant No. JPMXP1020230406) and JICFuS.
\end{acknowledgements}

\bibliography{ref, ref2}

\end{document}